\documentstyle[aps,floats,twocolumn,prd,psfig]{revtex}

\newcommand{\ApJL}{Astrophys. J. Lett.}
\newcommand{\ApJ}{Astrophys. J.} 
\newcommand{\ApJS}{Astrophys. J. Supp.} 
\newcommand{\PRL}{Phys. Rev. Lett.}
\newcommand{\PRD}{Phys. Rev. D}
\newcommand{\MNRAS}{Mon. Not. Roy. Astron. Soc.}
\newcommand{\AandA}{Astron. \& Astrop.}

\def\sun{\hbox{$\odot$}}

\newlength{\tskip}
\newlength{\colwidth}

\newcommand{\beq}{\begin{equation}}
\newcommand{\eeq}{\end{equation}}
\newcommand{\beqa}{\begin{eqnarray}}
\newcommand{\eeqa}{\end{eqnarray}}

\long\def\comment#1{}

\begin{document}   

\twocolumn[\hsize\textwidth\columnwidth\hsize\csname @twocolumnfalse\endcsname
\title{Graviton Mass from Close White Dwarf Binaries Detectable with LISA}
\author{Asantha Cooray and Naoki Seto }
\address{California Institute of Technology, Mail Code
130-33, Pasadena, CA 91125\\
E-mail: (asante,seto)@tapir.caltech.edu}

\maketitle
\begin{abstract}
The arrival times of gravitational waves and optical light from orbiting binaries provide
a mechanism to understand the propagation speed of gravity when compared to that of light or electromagnetic
radiation.
This is achieved with a measurement of any offset between 
optically derived orbital phase related to that derived from gravitational wave data,
at a specified location of one binary component with respect to the other.
Using a sample of close white dwarf binaries (CWDBs) detectable with the Laser Interferometer Space Antenna (LISA) and 
optical light curve data related to binary eclipses from meter-class telescopes for 
the same sample, we determine the accuracy to which orbital phase differences can be
 extracted. We consider an  application of these measurements 
involving a  variation to the speed of gravity, when compared to the speed of light, due to a massive graviton.
For a subsample of $\sim$ 400 CWDBs with high signal-to-noise gravitational wave and 
optical data with magnitudes brighter than 25, the combined upper limit on the graviton mass is at the
level of $\sim 6 \times 10^{-24}$ eV. This limit is two orders of magnitude better than the present limit derived by 
Yukawa-correction arguments related to the Newtonian potential and applied to the Solar-system.
\end{abstract}

\hfill
]

\section{Introduction}

In a recent paper, we discussed the optical followup study of close white dwarf binaries (CWDBs)
that will be detectable with the Laser Interferometer Space Antenna (LISA) mission \cite{Cooetal03}. 
Recent estimates suggest that close to 3000 binaries will be detected via gravitational waves (GWs) 
at frequencies above 3 mHz, with most of them
restricted to the frequency range between 3 mHz and 6 mHz \cite{webb,Neletal02,Set02}. These binaries
will only be localized to the accuracy of $\sim$ one sqr. degree with a three-year
LISA observation based on the Doppler effect related to the orbit 
around the Sun and the amplitude modulation due to the rotation of LISA detectors. 
For a precise location of an individual binary, optical observations are required; an
easy identification of the binary is facilitated by the presence of eclipses
in the optical light curve with a period twice the GW period.
Considering the impact on astronomy and astrophysics, in general, one can certainly expect
detailed optical followup observations of LISA binaries \cite{Cooetal03}. 
In addition to CWDB related physics, such as tidal heating, 
using both the optical light curve and gravitational wave data, we can also extract 
important information related to certain fundamental parameters in physics. We discuss such a possibility 
in the present paper by considering observational prospects in the LISA era.
 
In addition to a localization of the direction from which GWs
are emitted, with gravitational waves, LISA also allow a measurement of the binary distance,
the chirp mass, the binary orbital inclination angle, and the binary
orbital phase (modulus $\pi$).
The optical followup observations also allow a measurement of the orbital phase.
In general, one expects the orbital phase measured with gravity wave data to agree precisely
with that of the optical light curve. A difference in the two is only expected if 
optical light travel at a different speed than that of gravity, which determines the
propagation of gravitational waves. While a difference is not expected,
a potential non-zero mass related to the graviton particle can lead to a slight difference.
Thus, a comparison of the orbital phase related to an optical eclipsing light curve 
of CWDBs with the phase determined from LISA data will allow a constraint on the
graviton mass.

Note that previous studies have considered such a possibility, again with respect to 
LIGO and LISA data \cite{Wil98,Will:um,LarHis00,Cuetal02}.
While one of the considered methods involve the dispersion of the gravity waves alone, as a function
of the frequency \cite{Wil98,Will:um}, the comparison between optical or electromagnetic data and gravity waves is
expected to provide an improved constraint. Previous analyses, however, only considered a handful of
objects which are already known to be eclipsing binaries in optical data and
are expected to be GW sources detectable with LISA (e.g., 4U1820-30)
\cite{Cuetal02}. The present 
study considers a sample of CWDBs detectable and locatable
with LISA above its confusion noise and is bright enough optically for followup studies with
a few meter-class telescope.

The paper is organized as follows.  In \S~\ref{sec:lisa}, we
briefly discuss observations with LISA, requirements for an optical followup study and, then, consider
the extent to which orbital phases of individual binaries can
be determined. In \S~\ref{sec:results}, we
put these measurements in the context of an improved limit on the graviton mass.
We conclude with a summary in \S~\ref{sec:summary}.

\section{Close White Dwarf Binaries}
\label{sec:lisa}

First, we will review CWDB detections with LISA in gravitational waves 
and then move on to discuss aspects related to the optical light curve.
Our focus here would be related to the phase measurement related to the binary orbit, while
issues related to the CWDB localization with LISA data
and optical followup observations are discussed in Ref.~\cite{Cooetal03}, following initial calculations in 
Refs.~\cite{Cut98,TakSet02,Cornish:2003vj}.

\subsection{LISA Observations}

Briefly, with gravitational waves, one observes two components given 
by the quadrupole approximation in the principle polarization
coordinate \cite{Pet64}
\begin{eqnarray}
h_{+}(t) &=& A \cos \left[ 2\pi \left(f+\frac{1}{2} \dot{f} t \right) t + \varphi(t) \right] \times \left[1+ (\hat L \cdot \hat n)^2 \right] \nonumber \\
h_{\times}(t) &=& -2A \sin \left[ 2\pi \left(f+\frac{1}{2} \dot{f} t \right)t + \varphi(t)\right] \times (\hat L \cdot \hat n) \nonumber \\
\end{eqnarray}
where $\hat n$, given by $(\theta_S,\phi_S)$,  and $\hat L$ are unit directional vectors  
to the binary, from the observer, and the angular momentum of the binary, respectively. Here,
$\varphi(t)$ is the phase resulting from
the Doppler phase due to the revolution of LISA around the Sun:
\begin{equation}
\varphi(t) = 2 \pi f R \sin \theta_S \cos \left[ \frac{2 \pi t}{1 {\rm yr}} - \phi_S \right] + \varphi_0\, ,
\end{equation}
where  $R=1$ AU and $\varphi_0$ are integral constants.
The principle polarization coordinate is determined by two orthonormal
vectors  $\hat X$ and $\hat Y$ that are normal to the direction of the
source $\hat n$ through
\begin{eqnarray}
\hat X &=& \frac{\hat n \times \hat L}{|\hat n \times \hat L|}
 \nonumber \\ 
\hat Y &=& \hat n \times \hat X .
\end{eqnarray}

CWDBs are expected to have circular orbits due to the tidal interaction in
their early evolutional stage.  The amplitude of the wave is given by
\begin{equation}
A = \frac{5}{96 \pi^2} \frac{\dot{f}}{f^3 D} \, ,
\label{amplitude}
\end{equation}
where $D$ is the distance to the GW source.
Note that the GW frequency ($=2/P_{\rm orb}$ where $P_{\rm orb}$ is the
orbital period) for a circular orbit is related to the total mass
$M_1+M_2$ and the separation $a$
of the binary via
\begin{eqnarray}
f &=& 2 \sqrt{\frac{G(M_1+M_2)}{4\pi^2 a^3}} \nonumber \\
  &\approx&  3.5 \times 10^{-3} \left(\frac{M_{\rm tot}}{0.9 M_{\sun}}\right)^{1/2} \left(\frac{a}{10^5 {\rm km}}\right)^{-3/2} \; {\rm Hz}\, ,
\label{eqn:freq}
\end{eqnarray}
while the time variation of this frequency is
\begin{eqnarray}
&&\dot{f} = \frac{96 \pi^{8/3}}{5} {\tilde M}^{5/3}  f^{11/3} \nonumber \\
&\approx& 1.2 \times 10^{-16} \left(\frac{\tilde M}{0.4 M_{\sun}}\right)^{5/3} \left(\frac{M_{\rm tot}}{0.9 M_{\sun}}\right)^{11/6} \left(\frac{a}{10^5 {\rm km}}\right)^{-11/2}
\label{chirp}
\end{eqnarray}
when the chirp mass is given by ${\tilde M} = M_1^{3/5} M_2^{3/5} / (M_1+M_2)^{1/5}$ and the total mass $M_{\rm tot}=M_1+M_2$.

With gravitational waves, one can estimate a total of 8 independent quantities: $A$, $f$, $\dot{f}$, $\varphi_0$, location ($\vec n$, 2 parameters),
and the direction of the angular momentum, $\vec L$ (2 parameters).
At higher GW frequencies, one can also extract meaningful information on the
second derivative of the frequency $\ddot{f}$. The orbital inclination 
angle is given by $\cos^{-1} (\vec n \cdot \vec L)$.
In terms of physical quantities of interest, with $A$, $f$, $\dot{f}$,
one extracts $D$ and ${\tilde M}$; in addition to these parameters, one can also constrain a combination of
$M_{\rm tot}$ and $a$ with
Eq.~\ref{eqn:freq}. Note that relations (\ref{amplitude}) and
(\ref{chirp}) are given for Newtonian point particle systems as
the correction related to the finite size of the binary is not significant.

With parameters related to $\hat n$ and $\hat L$, one can reconstruct the 
orbital phase, say at a certain time $t$, which we define as
\begin{equation}
\phi(t) \equiv \tan^{-1} \left[ \frac{\hat r(t) \cdot \hat X}{\hat r(t) \cdot \hat Z}\right],
\end{equation}
where $\hat Z = \hat L \times \hat X$ and 
$\hat r(t)$ is the unit vector between the two masses. This vector changes as a function of time
as the binary components orbit with respect to  a common center of mass.

For the present discussion, we extend the calculation in Ref. \cite{Cooetal03} to
study how well the orbital phase in each of LISA detectable CWDBs can be determined. 
In order to generate the necessary representation of LISA observations, we
make use of the sample studied in \cite{Cooetal03} based on population synthesis code used by Ref.
\cite{FarPhi03}. This binary sample is distributed in the galaxy following
\begin{equation}
\rho(R,z) = \rho_0 {\rm sech}\left(\frac{|z|}{z_0}\right)^2 \exp\left(-\frac{R}{ R_0}\right) \, ,
\end{equation}
where $z_0 = 200$ pc and $R_0=2.5$ kpc \cite{Neletal01}.  For the sample of $\sim$ 
3000 binaries, we now randomly assign an orientation $\vec L $. This
means that the  inclination $\cos i$ is distributed uniformly in
$ 0 \leq \cos i \leq 1$.  For an optical eclipse to be observable, we require that the cosine of the
inclination angle be less than a minimum such that $\cos
i \le \cos i_{\rm min} \equiv (R_1+R_2)/a$.
For simplicity, here, we assume that binaries are equal mass. 
For a given  orbital frequency and a chirp mass, we calculate the ratio $
(R_1+R_2)/a$ as a function of $q\equiv M_1/M_{\rm tot}$ using 
a typical mass-radius relation of white dwarfs given in Ref.~\cite{nauenberg72}. 
We found that the ratio $
(R_1+R_2)/a$ takes a minimum value at $q=0.5$ and increases only $\sim
10\%$ at $q=0.3$ or $q=0.7$. This fractional change depends very weakly
on the chirp mass.   Considering the facts that the mass ratio
$M_1/M_2$ for  a CWDB  is expected to be close to 1 \cite{Maxted:2002yc},
our equal-mass assumption is conservative and safe.

Under these parameters, 
we extracted a subsample of objects from the total original sample that is expected to emit detectable GWs with
LISA and also show eclipses in the optical light curve.
Since $a$ typically  is few times $(R_1+R_2)$, the probability to observe an eclipse is rather high
and this subsample amounts to about 1400 objects with an overall probability of $\sim$ 0.45. The subsample size
is likely to vary higher by up to $\sim100$ objects when
 our assumption related to equal mass binaries is dropped.

The distribution of phase errors for this sample of $\sim$ 3000 CWDBs detectable with LISA is shown in Fig.~1. 
The phase error is calculated using 
the Fisher matrix approach discussed in Ref.~\cite{TakSet02}, where all 8 parameters related to
these observations are derived simultaneously from GW data. 
As in Ref.~\cite{TakSet02}, we use expressions 
related to LISA noise appropriate for the long wavelength limit, when compared to
arm-length, with a transfer function for the finite size of the arm-length that subsequently corrects 
for the first assumption.
This prescription is very effective for nearly monochromatic sources \cite{Seto:2002uj}. The noise curve  in
Ref. \cite{Finn:2000sy} is used here.  We consider three cases
with a total observational duration with LISA of 1, 3 and 5 years. The 
lower set of lines show the phase error related to the 
subsample that is expected to show eclipses in the optical light curve, simply selected based on the inclination angle. 
This subsample can be optically identified and localized precisely; however, assuming that
the location is exactly known is not important when the observational duration is 
more than 2 years (see, e.g., Ref.\cite{TakSet02}
for details on parameter improvement when the binary location is assumed to be known precisely a priori).

As discussed in Ref.~\cite{Cuetal02}, the phase error related to GWs is simply given by
\begin{equation}
\delta \phi_{\rm GW} = \frac{\alpha}{2} \left( \frac{{\rm S}}{{\rm N}}\right)^{-1} \left[ 1 +  {\cal O} (S/N)^{-1}\right]
\end{equation}
where, $\alpha$ is order unity or more and 
accounts for the fact that more than one parameter is to be determined from 
GW data.  The factor of $2$ in above accounts for the fact that we are 
determining the phase error related to the binary orbit,
which is  $1/2$ the gravity-wave phase.
The phase errors are generally at the level of $\sim$ 0.1, when the observational duration is more than 3 years. 
This is substantially
smaller than what was previously suggested in the literature (c.f. \cite{LarHis00})\footnote{The phase error
is independent of the timing error and the extent to which the period is known, as assumed in 
estimates by \cite{LarHis00}.
This can be simply understood by Fourier decomposing the light curve. The phase is simply related to the 
angle between that
of a Fourier mode and an arbitrary vector, related to the location from which the phase is measured, and 
the extent to which this angle can be determined depends on the length of the Fourier mode, which in return is
determined by the signal-to-noise of the light curve and not the timing accuracy.} 

In Fig.~1, for reference, we also show the phase error related to the nearby CWDB sample 
(at distances below 8 kpc). There are roughly
$\sim 1000$ CWDBs and we expect roughly $\sim$ 500 of them to show eclipses that can be detected in the optical 
light curve.
The phase error related to the nearby sample is better than the whole sample and results from the 
fact that the signal-to-noise for an individual detection is higher given the close distance. 

Note that the orbital phase determination is substantially improved with observational 
durations, $T_{\rm obs}$, at the level of 3 years.
While one generally expects the phase error determination to scale 
as $\sqrt{T_{\rm obs}}$, there is a substantial improvement
between one and three-year durations beyond that implied by this simple
scaling \cite{TakSet02}. This increase results from the fact that
significant degeneracies, between the binary location and other
parameters, are broken with observational durations 
greater than 2 years. We illustrate this in Fig.~2 where we show the
phase error distributions for parameter determination that involve (i)
all eight parameters, (ii) six parameters with location assumed to be known, 
and (iii) four parameters with both location and orbital orientation
assumed to be known. The 1 year curves with cases (i) and (ii) show the extent to which  degeneracies affect the
orbital phase determination.  With location assumed to be known, for an observational duration of 1
year, the degeneracy is broken and a substantial improvement in the
orbital phase determination is achieved.  In contrast, curves for cases (i) and
(ii) almost overlap when $T_{obs}=$3yrs or 5yrs. These results are
instructive to understand how the impact of precise localization (through
optical identification) on parameter estimation depends with the
observational duration of LISA.

Considering the observational situation relevant for this paper, we can
implicitly assume that individual binary locations are determined
precisely by the identification of optical counterparts. However, 
when we consider observational durations of order 3 years or more for GWs, 
the final graviton mass limit we calculate in the present paper
is essentially independent of whether the binary 
location is assumed to be known or not in the GW phase measurement.
One can obtain an improvement of a factor of  2 in the GW orbital phase when the orientation of the binary is also 
assumed to be known a priori.
We do not pursue this possibility, however, 
since this involves detailed modeling of the optical light curve or obtaining additional data at different
wavelengths of the electromagnetic radiation spectrum. For example,
we might get limited information on the inclination angle by modeling the
eclipse shape, but the polarization angle would not be determined.

\subsection{Optical data}

The optical magnitude distribution of the CWDB binary sample that is expected to be localized 
with LISA is discussed in Ref.~\cite{Cooetal03}. 
Most optical magnitudes are at the level of 25 and fainter.
Thus, we will only restrict the discussion  to the subsample that is bright. Similarly, the
extent to which the phase angle can be determined 
from the optical light curve depends on the single-to-noise ratio of the optical light curve,
and we write
\begin{equation}
\delta \phi_{\rm opt} = \beta \left( \frac{{\rm S}}{{\rm N}}\right)^{-1} \left[ 1 +  {\cal O} (S/N)^{-1}\right] \, ,
\end{equation}
where now $\beta$ is again a parameter of order unity or more as
we are trying to determine additional parameters from the optical data as well. Under the assumption that
the period is a priori known, from the GW data, and the assumption that the time derivative 
of the period is zero, which is a very safe assumption
since $\dot {P} \sim 10^{-10} s s^{-1}$ for 
binaries at 3 mHz with chirp mass $\sim 0.45$ M$_{\sun}$, $\beta$ is generally
at the level of 1 or slightly above.
Since the observational duration of GWs, $T_{{\rm obs,GW}} \sim 3$ years or more, is larger than the duration of optical observations ($\sim$ 100 minutes),  our assumption of
neglecting $\dot{P}$ for optical analysis if safe, though, we include $\dot{P}$, through $\dot{f}$, in estimating parameters related to GWs.

For a meter-class telescope, with 
an integration time of $t_{\rm int}$, the expected signal-to-noise ratio can be written as
\begin{equation}
\frac{\rm S}{\rm N} =  \frac{S_{\rm WD} t_{\rm int}}{\sqrt{S_{\rm WD} t_{\rm int} + S_{\rm sky} t_{\rm int} + S_{\rm det} t_{\rm int}}} \, ,
\end{equation}
where $S_{\rm WD}$, $S_{\rm sky}$ and $S_{\rm det}$ are the number of source, sky background and detector photons, respectively, per unit-time.
For a star of magnitude $m$, imaged with a telescope of diameter $D$, efficiency $\epsilon$, the number of photons, in a second, is
\begin{equation}
{\rm S}_{\rm WD} = 7.5 \times 10^2 \left(\frac{\epsilon}{0.5}\right) \left(\frac{10^{-2/5 m}}{10^{-10}}\right)\Big|_{m=20} \left(\frac{D}{4 {\rm m}}\right)^2  \, .
\end{equation}
In the limit that sky and detector backgrounds are not important, the signal-to-noise, for a 25-second integration is
\begin{equation}
\frac{\rm S}{\rm N} = 120 \left(\frac{\epsilon}{0.5}\right)^{1/2} \left(\frac{10^{-1/5 m}}{10^{-5}}\right)\Big|_{m=20} \left(\frac{D}{4 {\rm m}}\right) \left(\frac{t_{\rm int}}{25 {\rm sec}}\right)^{1/2} \, .
\end{equation}
In reality, at these faint magnitudes, the signal-to-noise  ratio is, however,
not determined only by the flux from the object, but rather by the sky and the detector
background, and the true signal-to-noise, for a 25 second observation, 
is at the level of 76 for $m=20$, 12 for $m=22.5$ and $\sim$ 1.5 for $m=25$, in the case of the 4-m Cerro-Tololo 
and Kitt Peak telescopes of the National Optical Astronomical Observatories\footnote{http://www.noao.edu/cgi-bin/scope/runiraf/ccdtime}.
In our calculations, we use these numbers, but restrict our attention to the subsample at a distance below 8 kpc and with magnitudes
brighter than $25$; with this additional restriction, the subsample reduces to about $\sim$ 400 CWDBs, for which reliable phase measurements may be possible from both GWs, with LISA, and optical data.

Given the magnitude distribution of the CWDBs that are expected to show optical eclipses, we can estimate the
expected distribution of errors in the orbital phase, determined from electromagnetic radiation. 
We summarize our results in Fig. 3 for the subsample of CWDBs, where we
also include extinction \cite{BahSon80}, following the modeling in Ref.~\cite{Cooetal03}. Most 
magnitudes are at the level of 25 and below, though for the sample with distances below
8 kpc, magnitudes are at the level of 22, 
albeit a broad distribution over the range of 20 to 25. While magnitudes here are
simply calculated based  on white dwarf cooling, given the age distribution determined in Ref. \cite{Cooetal03},
there is an additional effect that can potentially brighten CWDBs due to tidal heating by one white dwarf on the other
due to their close separation \cite{iben}. The magnitude distribution
gains few magnitudes as shown in Fig. 3, but we do not consider this possibility given the
complex physics associated with tidal heating 
such as the possibility that tidal heating is not uniform and will only lead to hot-spots among others.
However, we note  that, from detailed followup observation of
nearby CWDBs,  one can obtain information that would potentially
 affect the orbital phase measurement from optical data, {\it e.g.}  brightness
distribution of the surface of white dwarf. Note also that interacting
binaries, such as, AM CVn systems would be more luminous, but they
generally have smaller chirp masses (weaker gravitational wave signals),
and estimation of their orbital phases from optical data would not be as
simple as CWDBs (see e.g. \cite{LarHis00}).

\begin{figure}[!h]
\centerline{\psfig{file=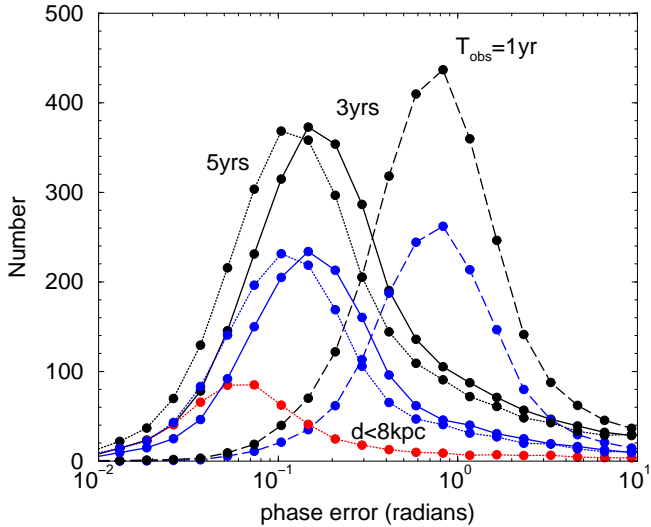,width=3.4in,angle=-90}}
\caption{The phase error distribution for GWs detectable with LISA for a sample $\sim$ 3000 CWDBs, and assuming a total observational
duration for 1, 3 and 5 years. The error in orbital phase is calculated following the Fisher matrix approach, with the total of eight 
parameters to be determine from gravity-wave data, including binary location. 
The lower curves are the error in orbital phase for the CWDBs that are
expected to show eclipses. For comparison, we also show phase errors related to the subsample, about $\sim$ 500 objects, that
are at distances below 8 kpc and are expected to show eclipses in the optical light curve, under the condition that $cos i \leq (R_1+R_2)/a$,
assuming a total observational duration of 5 years. On average, the phase measurement with the nearby sample is better since the expected
signal-to-noise for gravitational wave detection is higher.}
\label{fig:prim}
\end{figure}

\begin{figure}[!h]
\centerline{\psfig{file=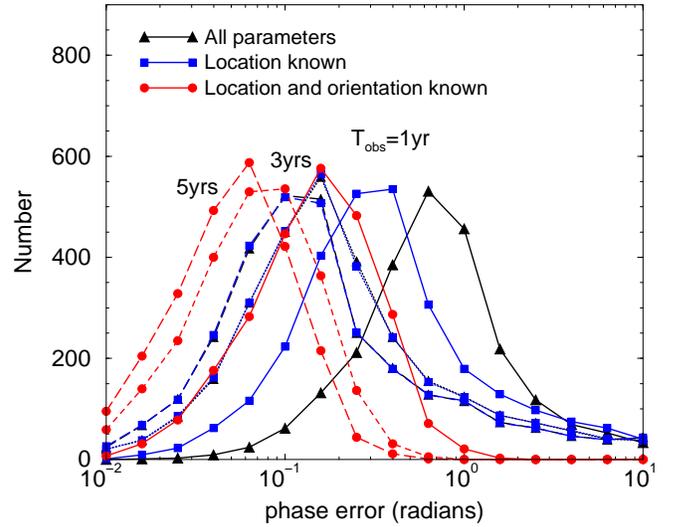,width=3.4in,angle=-90}}
\caption{We show the role of prior knowledge related to the binary location and orientation of when the orbital phase errors is determined from
GWs detectable with LISA for a sample $\sim$ 3000 CWDBs, and, again, assuming a total observational
duration of 1 (solid-lines), 3 (dotted-lines) and 5 (dashed-lines) years. The three sets of curves with triangles, squares and circles, represent the case with (i) no prior knowledge is available and all
 eight parameters are to be determined from the GW data, (ii) location assumed to be
known, and (iii) both location  and orientation of the binary is assumed to be known, respectively. In general, the location and
other parameters are highly correlated and precise localization leads to a substantial reduction in the accuracy of phase determination 
for observational durations of two years and less. For observational durations
greater than 3 years, the improvement in phase error related to a priori known location is not significant when compared to the case with all
parameters, including location, are to be determined from the
 data. Curves for cases (i) and (ii) are almost overlapped for
 $T_{obs}=$3yrs and 5yrs.  Additional degeneracies are broken when the binary orientation is also
assumed to be known and results in a  better determination of orbital phase. We do not pursue this possibility since
it involves detailed modeling of the optical light curve.}
\label{fig:gwphase}
\end{figure}

\begin{figure}[!h]
\centerline{\psfig{file=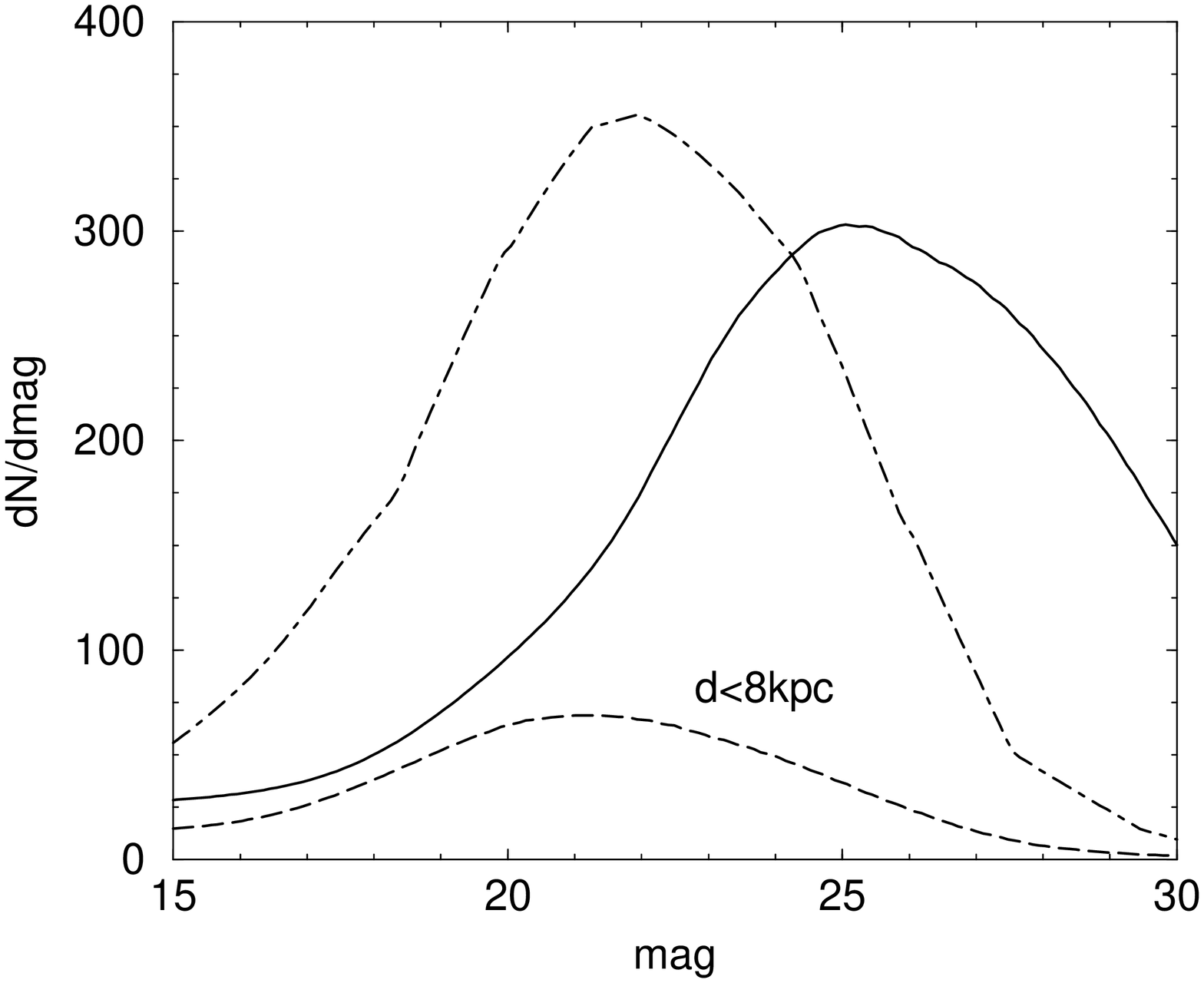,width=3.4in,angle=-90}}
\caption{The V-band magnitude distribution for the CWDB sample detectable with LISA and for which phase measurements may be possible. We
have taken into account the spatial distribution in addition to extinction (solid line). The long-dashed line show the
subsample at distances below 8 kpc. This subsample, on average, is brighter given the nearby distance and relatively low extinction, though the
distribution is some what broad. The dash-dotted line shows the possible
 effects of the tidal heating with using the extreme model in [20].}
\label{fig:mags}
\end{figure}

\section{Graviton Mass}
\label{sec:results}

With phase errors related to the GW and optical light curve data, one can consider a combined study involving the propagation speeds
of gravity ($c_g$) and light ($c$). The first obvious application is to determined if $c_g =c$ and to study any departures from it.
Compared to many other previous tests, this is an ideal 
scenario since one is comparing directly the gravitational radiation and with optical light
with a calibrated zero-point with respect to a certain location of the orbit.
The presence or the lack of any departure in the optically derived orbital phase from that of GWs
 can then be used to constrain some aspects related to the propagation that potentially
make two speeds depart from each other.  

As such a possibility, we consider the presence of a massive graviton such that the speed of gravity is modified
\begin{eqnarray}
c_g &=& c  \left[1 - \frac{m_g^2 c^4}{E^2}\right]^{1/2} \nonumber \\
    &\approx& c \left[1 - 1/2\frac{m_g^2 c^4}{E^2}\right] \, .
\label{eqn:cg}
\end{eqnarray}
For the orbital binary, assuming a gravity wave frequency of $f (= 2/{\rm P_{\rm orb}})$, and at a distance $D$
the two phases are
\begin{eqnarray}
\phi_{\rm opt} &=& 2 \frac{f D}{c} \\
\phi_{\rm GW} &=& \frac{f D}{c_g} \, . 
\end{eqnarray}
Simplifying with Eqs.~\ref{eqn:cg} and defining $\Delta \equiv \phi_{\rm GW}-\phi_{\rm opt}$, 
one leads to the constraint that 
\begin{equation}
m_gc^2 = h f \left[\frac{1}{2}\left(1+\frac{\pi f D}{c\Delta}\right)\right]^{-1/2} \, .
\end{equation}
In the absence of a measurable phase difference $\Delta=0$, an upper limit on $m_g$ can be
obtained with the limit on $\Delta$ such that
\begin{equation}
\Delta = \sqrt{\delta \phi_{\rm GW}^2 + \delta \phi_{\rm opt}^2} \, .
\end{equation}

\begin{figure}[!h]
\centerline{\psfig{file=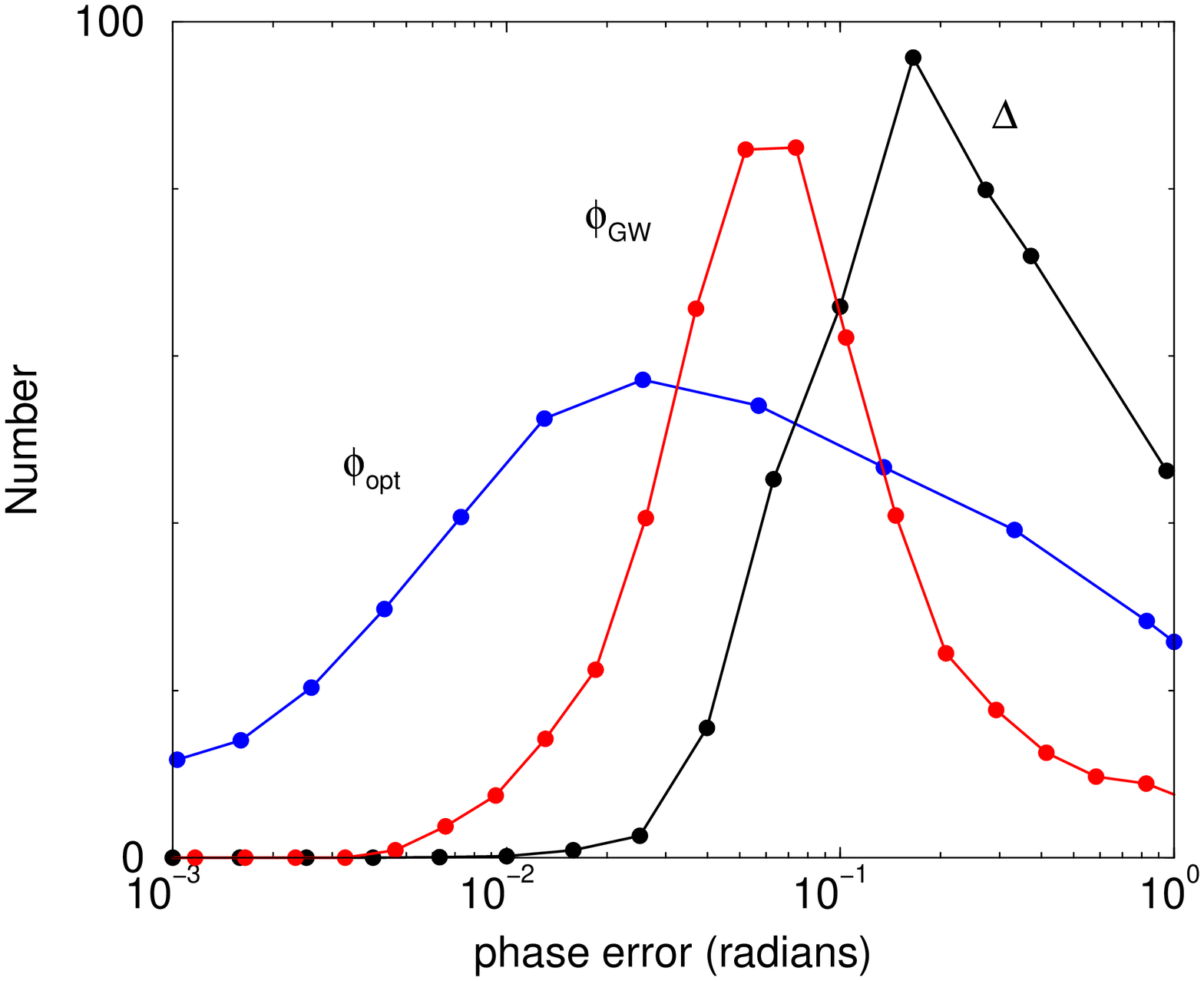,width=3.4in,angle=-90}}
\centerline{\psfig{file=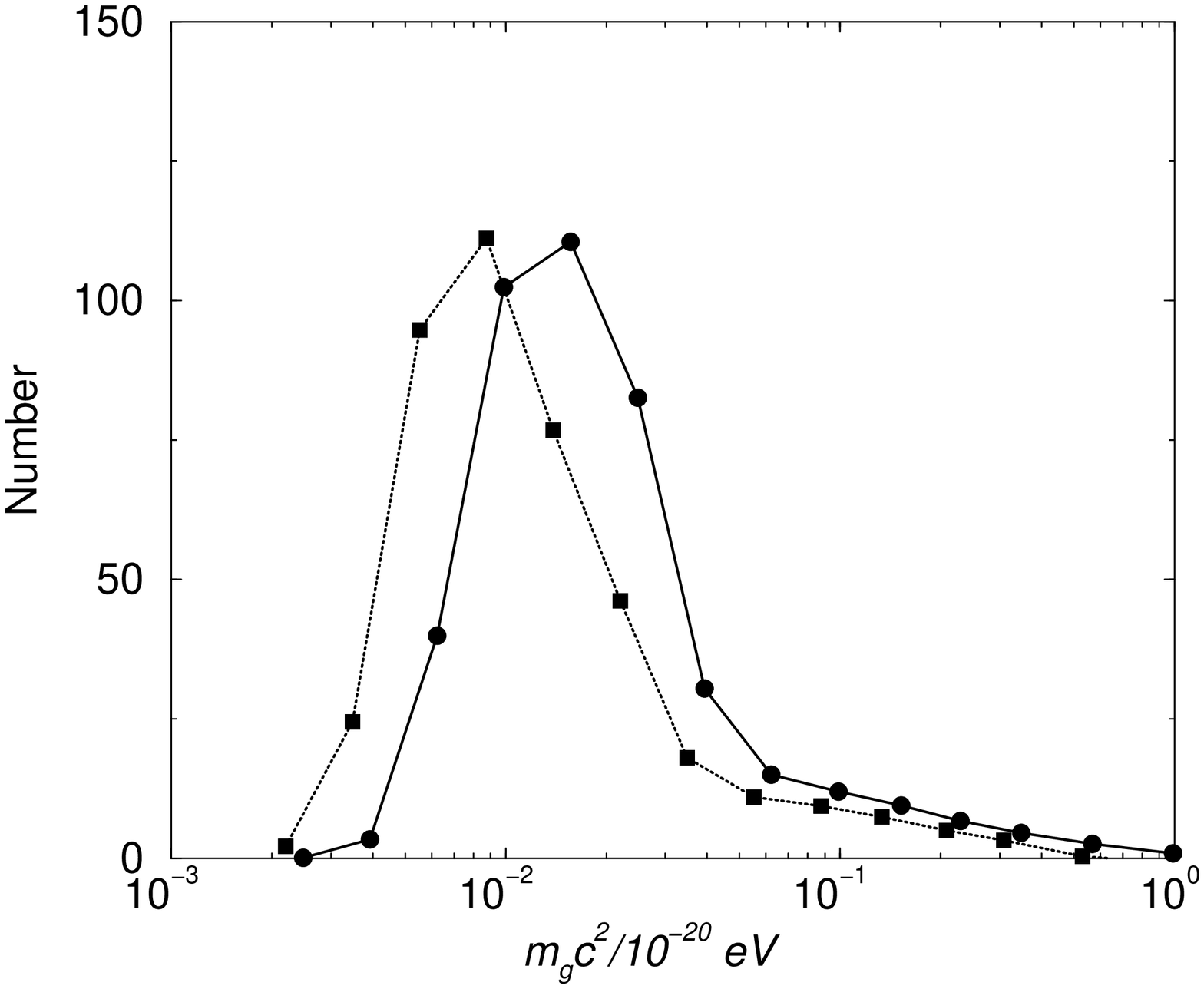,width=3.4in,angle=-90}}
\caption{{\it Top}: The error in orbital phase distribution for a subsample of $\sim$ 400 CWDBs at distances below 8 kpc and
at magnitudes brighter than 25.  We show separately the orbital phase error from GW and optical data and the combined error
related to the measurement of the phase difference in each of these
 cases. For each binary, we assume optical data of over 10 orbital cycles  or roughly over a time interval of
$\sim 6000$seconds, and consider the signal-to-noise measures as expected from a four-meter class telescopes (see, equations
11 to 13 and the discussion below Eq.~13). For the whole sample of 400 CWDBs, such a monitoring project takes
a total of $\sim$ 90 days assuming 8 hours of useful observational time each day.  {\it Bottom:} the distribution of graviton
mass limits derived from the above sample, together with its distance and frequency distributions. The limit, for an
individual CWDB, is generally at the level of 10$^{-22}$ eV, but for the sample as a whole, this limit improves to
about $\sim 6 \times 10^{-24}$ eV. The dotted line is the results only
 with considering the phase errors of gravitational waves. The combined
 limit becomes $\sim 3\times 10^{-24}$ eV.}
\label{fig:limit}
\end{figure}

In the limit that $\delta \phi_{\rm GW} >> \delta \phi_{\rm opt}$, the limit on graviton mass
can be written in terms of the signal-to-noise to which LISA detects gravitational waves \cite{Cuetal02}
\begin{equation}
m_g c^2 < \frac{h \sqrt{\alpha}}{\pi} \sqrt{\frac{ \pi f c}{D}} \left(\frac{{\rm S}}{{\rm N}}\right)^{-1/2} \, ,
\end{equation}
and since the GW signal-to-noise simply scales as $1/D$, this limit is independent on the distance at which the binary
is located \cite{Cuetal02}. In contrast the signal to noise ratio of electro-magnetic
waves scales as $1/D^2$, such that in the opposite limit, the limit on graviton mass is no longer independent of the distance. 

As shown in Fig.~1, and implied from Fig.~3, either one of these assumptions, $\delta \phi_{\rm GW} >> \delta \phi_{\rm opt}$ or
$\delta \phi_{\rm GW} << \delta \phi_{\rm opt}$, however, is not correct since the error to which both
optical and gravity-wave phases are determined is at the same order.
Thus, one should include both errors in 
estimating the limit on graviton mass. For the present discussion, we make use of the sample at a distance
below 8 kpc.  The subsample is brighter and is within the magnitude limits comfortably reachable with few meter or more-class
telescopes that are currently available. We make use of a fraction of 400 objects with magnitudes brighter than 25, and with distances
below 8 kpc. Here, we neglect the uncertainty in the  distance estimation related to LISA data since its error is in the 
second order. The individual distances are, however,  statistically included by accounting for the  distance distribution
function based on our model assumption related to the spatial distribution following Eq.~8.

Note that the observations we consider are independent of the localization of CWDBs since this can be achieved at a 
high signal-to-noise ratio by imaging at time intervals corresponding to expected minima
and maxima in the optical light curve based on prior phase information from the gravity-wave data. To obtain the optical
phase precisely, detailed sampling of the light curve is required and will involve imaging at time intervals of
order 25 seconds, given that the periods are mostly between 400 to 800 seconds.
The orbital phase errors related to optical and GW data are summarized in Fig.~4 for the subsample of CWDBs.
The phase errors are generally at the level of 10$^{-1}$ for GWs, while the orbital phase error from the optical data follow
a broad distribution. The error related to the combined phase difference is also shown in Fig.~4, which is calculated using 
physical properties of the subsample. 
In general, note that the phase error distribution related to GWs and optical data are not independent;
the CWDBs with better determined orbital phases with GW data are expected to be nearby and, thus, to be bright optically;
For these nearby ones, orbital phases from  optical light curve data are
also better determined.  

In Fig.~4 bottom plot, we summarize the extent to which the graviton mass can be constrained from each of the CWDBs
at distances below 8 kpc and with magnitudes better than 25. The one-sigma limit on the graviton mass is generally at
the level of 10$^{-22}$ eV. If phase information related to the whole sample can be combined for a single constraint
as whole, the combined limit, by taking $(\sum_i 1/\sigma_i^2)^{-1/2}$ improves to about $6 \times 10^{-24}$ eV at the one-sigma level.  This can be compared to
current limits on the graviton mass.  Based on the Yukawa-correction to a
Newtonian potential, in the presence of a massive graviton,
the limit from Solar-system dynamics is $4.4 \times 10^{-22}$ eV \cite{Taletal88}, though the limit improves substantially
to $2 \times 10^{-29}$ eV with galaxy clusters at Mpc size-scales \cite{GolNie74,Har73}. The 90\% confidence limit from the orbital
decay of two-binary pulsars, again an indirect estimate on the graviton mass, is $7.6 \times 10^{-20}$ eV \cite{SutFin02}.
The direct limits from GW and optical data are discussed in Refs. \cite{LarHis00} and \cite{Cuetal02}; 
these limits are generally above the
level we have discussed, as they usually involve either a single binary or different assumptions and estimates.
Our limit, for the sample as a whole, is consistent with the best possible limit from LISA data implied in Ref.~\cite{Cuetal02}.

In general, we believe that the CWDB sample related to LISA provides a well controlled sample with
adequate optical followup opportunities to obtain a reliable limit on the graviton mass
as a whole. When obtained, such a limit will be independent of indirect assumptions and will be directly based on the propagation speeds
of gravity and light.
Finally we comment on a potential contaminant related to this measurement involving  gravitational lensing 
of the background binaries by the foreground mass distribution. Note that the lensing
probability in our galaxy has been measured to be at the level of  $\tau\sim 10^{-6}$ \cite{Alc00}, suggesting that
the probability to observe a lensed CWDB is substantially small. 
Since gravitational waves from  CWDBs have much lower frequencies $(\sim 10^{-2}{\rm Hz})$ 
 than optical  waves $\sim 10^{16}$Hz, the critical frequency  for lensing with a mass $M_{lens}$ is
$f_{cr}\sim(GM_{lens}/c^3)^{-1}\sim  10^{4} (M_{lens}/M_\odot)^{-1}$Hz.  For example, 
the amplification effect is almost negligible
for waves with $f\ll f_{cr}$ where the geometrical optics approximation
is not  valid \cite{Nakamura:1997sw,Takahashi:2003ix,Yamamoto:2003cd}. There
might be effective time delay between the gravitational wave and the long wave limit
(electro-magnetic waves)  at most $GM_{lens}/c^3\sim 10^{-4}$seconds with 
a phase shift at the level of $\sim 10^{-6}$  radians. This, however,  is much smaller than our resolution
$\Delta \phi/(2\pi f)\sim 10$ seconds and well below the phase measurement error. Given the low probability for lensing
and the small phase difference expected, we do not consider lensing contamination to be a major source of concern.

Other potential contaminants include modifications to the speed of light, relative to gravity waves,
based on refractive fluctuations through the neutral interstellar medium, such as gas clouds. Gravity waves
will pass such clouds unobscured, while density gradients within the medium will lead to refraction
and, thus, modifications to the path length of the optical light.
Here, again, expectations are relatively small modifications, if any, but we suggest that further
studies may be necessary given the importance of combined GW and optical studies in the LISA era and the unique
sample of CWDBs that LISA data are expected to provide.
  
\section{Summary}
\label{sec:summary}

The arrival times of gravitational waves and optical light from orbiting binaries provide
a mechanism to understand the propagation effects of gravity, when compared to light.
This is easily achieved via binary orbital phase measurement and by looking for an phase offset
in the optically derived orbital phase related to that derived from gravity wave data.
Using a sample of close white dwarf binaries (CWDBs) detectable with the Laser Interferometer Space Antenna (LISA) and the
associated optical light curve data related to binary eclipses, we determine the accuracy to
which orbital phase difference can be extracted. We consider an
application of these measurements, determining   an upper limit on the graviton mass.

For a subsample of $\sim$ 400 CWDBs with high signal-to-noise gravity wave and optical data with magnitudes brighter than 25,
the combined upper limit on the graviton mass is at the
level of $\sim 6 \times 10^{-24}$ eV, which is two orders of magnitude better than the limit derived by
Yukawa-correction arguments related to the Newtonian potential and applied to the Solar-system.

\acknowledgments
We thank Alison Farmer for information related to the binary sample and Daniel Holz for pointing out
previous work on this subject. This work was supported in part by DoE
DE-FG03-92-ER40701, 
a senior research fellowship from the Sherman Fairchild foundation (AC)
and NASA grant NAG5-10707 (NS).


\begin{thebibliography}{99}
\frenchspacing


\bibitem{Cooetal03}
 A. Cooray, A. Farmer and N. Seto,
 \ApJL, {\bf 601},
 L47 (2004).

\bibitem{webb}
R. F. Webbink and Z. Han,  AIP Conf. Proc. 456: Laser Interferometer
Space Antenna, Second International LISA Symposium on the
Detection and Observation of Gravitational Waves in Space. (AIP,
New York, 1998) p. 61.

\bibitem{Neletal02}
  G. Nelemans, L. R. Yungelson, and S. F. Portegies Zwart  \AandA, {\bf 375}, 890 (2001).

\bibitem{Set02}
N.~Seto,
Mon.\ Not.\ Roy.\ Astron.\ Soc.\  {\bf 333}, 469 (2002)

\bibitem{Wil98} C. M. Will, \PRD\ {\bf 57}, 2061 (1998).

\bibitem{Will:um}
C.~M.~Will,
Class.\ Quant.\ Grav.\  {\bf 20} (2003) S219.

\bibitem{LarHis00} S. L. Larson and W. A. Hiscock, \PRD\ {\bf 61}, 104008 (2000).

\bibitem{Cuetal02} C. Cutler, W. A. Hiscock and S. L. Larson, \PRD\ {\bf 67}, 024015 (2003).

\bibitem{Cut98} C. Cutler and E. Flanagan, \PRD\ {\bf 49}, 2658 (1994); C. Cutler, \PRD\ {\bf 57}, 7089 (1998).

\bibitem{Pet64}P. C. Peters, \PRD\ {\bf 136}, 1224 (1964).

\bibitem{FarPhi03} A. Farmer and E. S. Phinney, 
Mon.\ Not.\ Roy.\ Astron.\ Soc.\  {\bf 346}, 1197 (2003)
[arXiv:astro-ph/0304393].



\bibitem{Neletal01}
  G. Nelemans, L. R. Yungelson, S. F. Portegies Zwart and F. Verbunt, \AandA, {\bf 365}, 492 (2001).

\bibitem{TakSet02}
R.~Takahashi and N.~Seto,
Astrophys.\ J.\  {\bf 575}, 1030 (2002)



\bibitem{Cornish:2003vj}
N.~J.~Cornish and S.~L.~Larson,
Phys.\ Rev.\ D {\bf 67}, 103001 (2003).



\bibitem{nauenberg72} M. Nauenberg, \ApJ\ {\bf 175}, 417 (1972).

\bibitem{Maxted:2002yc}
P.~F.~Maxted, T.~R.~Marsh and C.~K.~Moran,
\MNRAS\ {\bf 332}, 745 (2002).




\bibitem{Seto:2002uj}
N.~Seto,
Phys.\ Rev.\ D {\bf 66}, 122001 (2002)



\bibitem{Finn:2000sy}
L.~S.~Finn and K.~S.~Thorne,
Phys.\ Rev.\ D {\bf 62}, 124021 (2000).


\bibitem{BahSon80}
J. N. Bahcall and R. M. Soneira, \ApJS\ {\bf 44}, 73 (1980).

\bibitem{iben}
I. Iben,  A. V. Tutukov   and A. V. Fedorova,  \ApJ\ {\bf 503}, 344 (1998).




\bibitem{Taletal88} C. Talmadge, J.-P. Berthias, R. W. Hellings and E. M. Standish, \PRL\ {\bf 61}, 1159 (1988).

\bibitem{GolNie74} A. S. Goldhaber and N. M. Nieto, \PRD\ {\bf 9}, 1119 (1974).

\bibitem{Har73} M. G. Hare,  Can. J. Phys. {\bf 51}, 431 (1973).

\bibitem{SutFin02} P. J. Sutton and L. S. Finn, Class. Quant. Grav. {\bf 19}, 1355 (2002).


\bibitem{Alc00} C. Alcock, R. A. Allsman, D. R. Alves et al. \ApJ\ {\bf
 541}, 734 (2000).


\bibitem{Nakamura:1997sw}
T.~T.~Nakamura,
Phys.\ Rev.\ Lett.\  {\bf 80}, 1138 (1998).


\bibitem{Takahashi:2003ix}
R.~Takahashi and T.~Nakamura,
Astrophys.\ J.\  {\bf 595}, 1039 (2003).



\bibitem{Yamamoto:2003cd}
K.~Yamamoto,
arXiv:astro-ph/0309696.



\end{thebibliography}
\end{document}